\documentclass[sigconf, authorversion]{acmart}

\AtBeginDocument{%
  }

\copyrightyear{2026}
\acmYear{2026}
\setcopyright{cc}
\setcctype{by}
\acmConference[ITiCSE 2026]{Proceedings of the 31st ACM Conference on Innovation and Technology in Computer Science Education V. 1}{July 10--15, 2026}{Madrid, Spain}
\acmBooktitle{Proceedings of the 31st ACM Conference on Innovation and Technology in Computer Science Education V. 1 (ITiCSE 2026), July 10--15, 2026, Madrid, Spain}
\acmDOI{10.1145/3803400.3809312}
\acmISBN{979-8-4007-2634-7/2026/07}

\usepackage[utf8]{inputenc} % allow utf-8 input
\usepackage{subcaption}
\usepackage[inline]{enumitem}

\usepackage{caption}
\usepackage{graphicx}
\usepackage{listings}

\newcommand{\rqbox}[1]{%
  \par\smallskip
  \noindent
  \setlength{\fboxsep}{4pt}%
  \fcolorbox{black!40}{gray!10}{%
    \parbox{\dimexpr\linewidth-2\fboxsep-2\fboxrule\relax}{#1}%
  }%
}

\begin{document}

\title[Interleaving Natural Language Prompting with Code Editing for Solving Programming Tasks]{Interleaving Natural Language Prompting with Code Editing \\ for Solving Programming Tasks with Generative AI Models}

\author{Victor-Alexandru P{\u a}durean}
\affiliation{
  \institution{MPI-SWS}
  \city{Saarbr{\"u}cken}
  \country{Germany}  
}
\email{vpadurea@mpi-sws.org}
\orcid{0009-0004-2998-096X}

\author{Alkis Gotovos}
\affiliation{
  \institution{MPI-SWS}
  \city{Saarbr{\"u}cken}
  \country{Germany}  
}
\email{agkotovo@mpi-sws.org}
\orcid{0000-0002-3902-8890}

\author{Ahana Ghosh}
\affiliation{
  \institution{MPI-SWS}
  \city{Saarbr{\"u}cken}
  \country{Germany}  
}
\email{gahana@mpi-sws.org}
\orcid{0000-0002-0967-5886}

\author{Paul Denny}
\affiliation{
  \institution{University of Auckland}
  \city{Auckland}
  \country{New Zealand}  
}
\email{paul@cs.auckland.ac.nz}
\orcid{0000-0002-5150-9806}

\author{Juho Leinonen}
\affiliation{
  \institution{Aalto University}
  \city{Espoo}
  \country{Finland}  
}
\email{juho.2.leinonen@aalto.fi}
\orcid{0000-0001-6829-9449}

\author{Andrew Luxton-Reilly}
\affiliation{
  \institution{University of Auckland}
  \city{Auckland}
  \country{New Zealand}  
}
\email{andrew@cs.auckland.ac.nz}
\orcid{0000-0001-8269-2909}

\author{James Prather}
\affiliation{
  \institution{Abilene Christian University}
  \city{Abilene}
  \state{TX}
  \country{USA}
}
\email{james.prather@acu.edu}
\orcid{0000-0003-2807-6042}

\author{Adish Singla}
\affiliation{
  \institution{MPI-SWS}
  \city{Saarbr{\"u}cken}
  \country{Germany}  
}
\email{adishs@mpi-sws.org}
\orcid{0000-0001-9922-0668}

\renewcommand{\shortauthors}{Victor-Alexandru P{\u a}durean et al.}
%% No italics, no superscripts, not anonymous
%% Use footnote or author note to identify equal contribution and/or contact author info

%%%%%%%%%%%%%%%%%%%%%%%%%%%%%%%%%%%%%%%%%%%% ABSTRACT
%%%%%%%%%%%%%%%%%%%%%%%%%%%%%%%%%%%%%%%%%%%%%%%%%%%%%%%%%%%%%%%%%%
%%%%%%%%%%%%%%%%%%%%%%%%%%%%%%%%%%%%%%%%%%%%%%%%%%%%%%%%%%%%%%%%%%
\begin{abstract}
Modern computing students often rely on both natural-language prompting and manual code editing to solve programming tasks. Yet we still lack a clear understanding of how these two modes are combined in practice, and how their usage varies with task complexity and student ability. In this paper, we investigate this through a large-scale study in an introductory programming course, collecting $13,305$ interactions from $355$ students during a three-day lab activity. Our analysis shows that students primarily use prompting to generate initial solutions, and then often enter short edit-run loops to refine their code following a failed execution. Student reflections confirm that prompting is helpful for structuring solutions, editing is effective for making targeted corrections, while both are useful for learning. We find that manual editing becomes more frequent as task complexity increases, but most edits remain concise, with many affecting a single line of code. Higher-performing students succeed with less reliance on editing and fewer overall interactions. These findings highlight the role of manual editing as a form of last-mile repair, complementing prompting in AI-assisted programming workflows.
\end{abstract}




\maketitle

% content
%%%%%%%%%%%%%%%%%%%%%%%%%%%%%%%%%%%%%%%%%%%% INTRODUCTION
%%%%%%%%%%%%%%%%%%%%%%%%%%%%%%%%%%%%%%%%%%%%%%%%%%%%%%%%%%%%%%%%%%
%%%%%%%%%%%%%%%%%%%%%%%%%%%%%%%%%%%%%%%%%%%%%%%%%%%%%%%%%%%%%%%%%%
\section{Introduction}\label{sec.introduction}

\looseness-1Imagine a student typing the following prompt to a generative AI (GenAI) model: \emph{``Write a function returning the position of the last occurrence of a positive number in a list''}. Syntactically correct Python code is generated almost instantly. The student scans the code, manually editing the function name to match the test suite. Upon execution, several test cases fail due to a mishandled boundary condition. Instead of re-prompting the model, the student manually edits a logical operator. They test the function again, this time successfully.

\begin{figure*}[t!]
    \centering
    \begin{subfigure}{0.40\textwidth}
        \centering
        \setlength{\fboxsep}{10pt}
        \fcolorbox{gray!70!black}{gray!10}{%
          \parbox{\dimexpr\linewidth-2\fboxsep-2\fboxrule\relax}{%
            \small
            In this exercise, you will design a function
    
            \bigskip
            \texttt{process\_list(lst, start, end)}
    
            \bigskip
            After carefully looking at the provided specifications, write your prompts to interact with the AI model and guide it to generate a correct program.
          }%
        }
    
        \par\vspace{13.75mm}
    
        \includegraphics[width=0.99\textwidth]{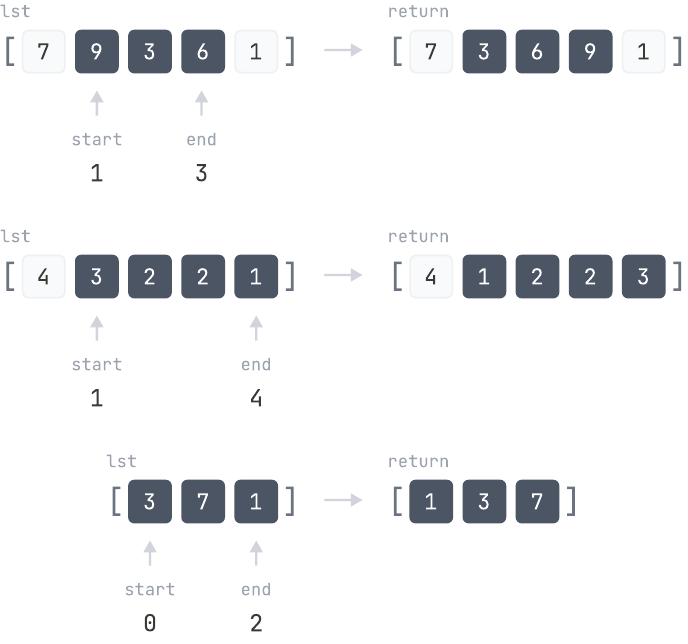}
    
        \par\vspace{13.75mm}
    
        \caption{Visual input-output specifications.}
        \vspace{-0.5mm}
        \label{fig.illustration.description}
    \end{subfigure}
    \hfill
    \begin{subfigure}{0.57\textwidth}
        \centering
        \includegraphics[width=\textwidth]{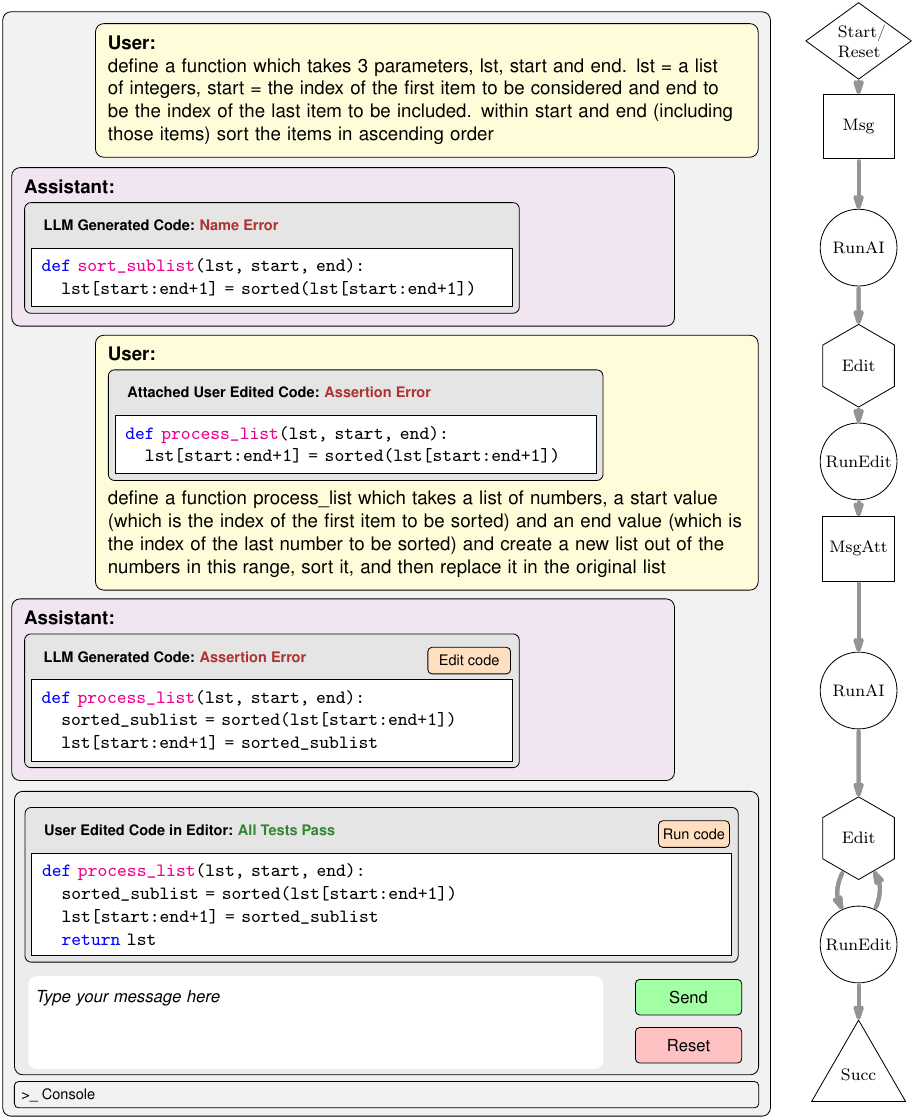}
        \caption{Example of a student's interaction combining NL prompting and code editing.}
        \vspace{-0.5mm}
        \label{fig.illustration.interaction}
    \end{subfigure}
    \caption{\looseness-1Illustration of a student's interaction combining prompting and editing. (a) presents the specification for the `sort sub-list' problem, where a sub-list must be sorted within the given index range. (b) recreates a genuine student's interaction, demonstrating how they interleave natural-language prompting and manual code editing, summarized by the state-transition diagram.}
    \Description[Two-panel figure showing a programming task and a student interaction]{Two-panel figure illustrating how a student combines prompting and manual code editing. The left panel shows the visual specification for a function named process_list with example input-output diagrams for sorting a sub-list within a given index range. The right panel shows a genuine chat-based interaction: the student prompts for code, receives a solution with a name error, edits the code, sends the edited version back with a revised prompt, receives another failing solution, and then goes on to make manual edits in the built-in editor, the last edit passing all tests. A small state-transition diagram summarizes the sequence Start/Reset, message, run generated code, edit, run edited code, message with attached code, run generated code again, edit, run edited code, and success.}
    \label{fig.illustration}
\end{figure*}

\looseness-1The type of interaction just described is becoming increasingly common in programming classrooms \cite{vadaparty2024cs1llm, kazemitabaar2024novices}.  For many students, writing a program no longer starts with a blank editor, but with code produced by a natural language prompt.  A well-crafted prompt can often produce code that serves as a good starting point, overcoming the complexities of syntax and cryptic compiler feedback that have traditionally been barriers for novices \cite{becker2019compiler, edwards2020syntax, denny2011understanding}.  However, even for simple programming tasks, students rarely craft prompts that are successful on the very first attempt \cite{padurean2025prompt}.  This raises the question of what students do next, and why. Do they revise their prompt, or dive into the code and make manual changes? This decision point is a critical, yet understudied, aspect of modern programming education: the role of manual code editing in an AI-assisted workflow. 

Exploring this choice is important because of the concern from some educators that students may over-rely on GenAI, avoiding mastery of programming fundamentals \cite{prather2024widening, cambaz2024use}. If problems appear solvable through prompting alone, students may fail to appreciate the precision and structure involved in writing and debugging code. This concern aligns with arguments highlighting limitations of natural language as a programming interface. Yellin emphasizes that natural language prompts are often insufficiently precise for specifying complex software behaviors~\cite{yellin2023premature}. Empirical evidence supports these limitations: in an upper-division software engineering course, Shah et al. found that students frequently needed to manually verify and debug code generated by GitHub Copilot~\cite{shah2025students}.

\looseness-1In this work, we explore how students integrate these two emerging competencies -- natural language prompting and manual code editing -- when solving programming problems. We build on the idea of Prompt Problems, visual computational tasks that students solve via natural language prompting \cite{denny2024prompt}. We use Prompt Programming \cite{padurean2025prompt}, a publicly available web platform where students can solve visually specified computational tasks through a combination of prompt-based code generation and direct code editing. Figure~\ref{fig.illustration} presents a genuine student conversation and its corresponding state diagram. The student begins a new conversation (Start/Reset) by describing the task in natural language (Msg), then runs the GenAI-generated code (RunAI), which fails. They switch to manual editing (Edit) and run the edited code (RunEdit), but a different issue comes up. To tackle it, they attach the buggy code to a new prompt (MsgAtt), receive revised code from the GenAI, and run it again (RunAI). When that also fails, they return to manual editing. The student then iterates through three cycles of editing and running code (loop between Edit and RunEdit; only the final code is shown in the figure), and ultimately reaches a successful solution (Succ). 

We collected such student interactions through the platform, and separately, asked students to reflect on their strategies after completing the task, describing what influenced their choice of when to prompt or edit and how they navigated the task. Using both these behavioral traces and student reflections, our analysis is guided by the following research questions (RQs):

\setlength{\leftmargini}{1em}
\begin{itemize} 
\item \textbf{RQ1}: How do students interleave natural-language prompting and manual code editing while solving programming tasks w.r.t. both behavioral patterns and perceived usefulness?
\item \textbf{RQ2}: How does task complexity influence students' code-editing behavior?
\item \textbf{RQ3}: How does students' programming competence influence their code-editing behavior?
\end{itemize}

%%%%%%%%%%%%%%%%%%%%%%%%%%%%%%%%%%%%%%%%%%%%%%%%%%%%%%%%%%%%%%%%%%
%%%%%%%%%%%%%%%%%%%%%%%%%%%%%%%%%%%%%%%%%%%%%%%%%%%%%%%%%%%%%%%%%%

%%%%%%%%%%%%%%%%%%%%%%%%%%%%%%%%%%%%%%%%%%%% RELATED-WORK
%%%%%%%%%%%%%%%%%%%%%%%%%%%%%%%%%%%%%%%%%%%%%%%%%%%%%%%%%%%%%%%%%%
%%%%%%%%%%%%%%%%%%%%%%%%%%%%%%%%%%%%%%%%%%%%%%%%%%%%%%%%%%%%%%%%%%
\section{Related Work}\label{sec.relatedwork}

\looseness-1\textbf{Generative AI Tools in Computing Education.} Generative AI is reshaping computing education~\cite{DBLP:journals/corr/abs-2402-01580,prather2023navigating}. Beyond code explanation generation \cite{macneil23sigcse,leinonen2023comparing}, buggy program repair \cite{DBLP:journals/pacmpl/ZhangCGLPSV24}, personalized feedback, and clearer error messages~\cite{DBLP:conf/sigcse/WangMP24,DBLP:conf/edm/PhungCGKMSS23}, these models increasingly help with generation of educational content, including quizzes, programming tasks, and debugging exercises \cite{padurean2024neural,doughty2024comparative,ma2024hypocompass,padurean2024bugspotter,nguyen2025synthesizing}. Furthermore, conversational agents based on GenAI enable interactive dialogues, simulating classroom scenarios and facilitating personalized tutoring across educational contexts \cite{DBLP:conf/lats/MarkelOLP23,DBLP:conf/edm/NguyenTS24,acun2024gaienhanced,zamfirescupereira2024conversational,DBLP:conf/aied/SchmuckerXAM24}. Despite these promising applications, research also highlights the need for careful management of AI-supported interactions to ensure that students remain engaged and can critically evaluate generated outputs, and genuinely develop computational thinking skills rather than relying passively on model-generated solutions \cite{kazemitabaar2024codeaid,denny2024desirable,prather2024widening,amoozadeh2024student}.

\textbf{Prompt Problems and Prompt Programming.} The concept of Prompt Problems, introduced by Denny et al.~\cite{denny2024prompt}, challenges students to solve computational tasks presented in a visual manner, by writing prompts for AI models. This approach has been shown to increase student engagement and lower syntactic barriers~\cite{denny2024prompt}. Further work built upon these tasks by supporting multilingual prompts for non-native English speakers~\cite{prather2024breaking} and introducing tasks where students craft prompts to generate code equivalent to provided examples~\cite{kerslake2024integrating}. Moreover, platforms such as Prompt Programming emerged as well, enabling iterative prompt refinements via chat-like interfaces and richer problem-solving interactions~\cite{padurean2025prompt}. However, these studies mainly investigated scenarios where problems can be successfully solved only through natural language prompting, leaving unexplored the critical moments when students must decide between further prompting or manually editing generated code.

\looseness-1\textbf{Natural Language Prompting vs. Code Editing.} Educators raise concerns that easy access to AI-generated code can encourage passive reliance rather than active evaluation~\cite{prather2024widening,DBLP:conf/chi/NguyenBZGAF24}. Prior work has shown that learners may struggle to assess and refine AI-generated solutions, often accepting initial outputs without reviewing or validating them~\cite{DBLP:journals/corr/abs-2501-10365,prather2024widening}. A similar line of research has shown that natural language can be too vague for describing exactly what a program should do as tasks get harder~\cite{yellin2023premature,huttel2024program}. Studies of tools like GitHub Copilot found that students often need to manually edit code after prompting, suggesting that prompting alone usually is not enough to solve many problems~\cite{shah2025students,vadaparty2024cs1llm}. However, detailed analyses investigating how students navigate prompting-to-editing transitions, and what influences their choices, remain limited.
%%%%%%%%%%%%%%%%%%%%%%%%%%%%%%%%%%%%%%%%%%%%%%%%%%%%%%%%%%%%%%%%%%
%%%%%%%%%%%%%%%%%%%%%%%%%%%%%%%%%%%%%%%%%%%%%%%%%%%%%%%%%%%%%%%%%%

%%%%%%%%%%%%%%%%%%%%%%%%%%%%%%%%%%%%%%%%%%%% METHOD
%%%%%%%%%%%%%%%%%%%%%%%%%%%%%%%%%%%%%%%%%%%%%%%%%%%%%%%%%%%%%%%%%%
%%%%%%%%%%%%%%%%%%%%%%%%%%%%%%%%%%%%%%%%%%%%%%%%%%%%%%%%%%%%%%%%%%

\section{Methodology}\label{sec.methodology}

This section outlines the platform used for our study, the classroom deployment, and descriptive statistics of the collected data.

\subsection{Tool Overview}

To conduct our study, we used the Prompt Programming platform \cite{padurean2025prompt}. The publicly available platform supports a chat-based workflow where students can solve programming problems by writing natural-language prompts, manually editing generated code, and running solutions against hidden tests. We selected our problems from the platform's existing problem set and analyzed anonymized interaction logs collected through the platform. We provide further details about the platform next.

\looseness-1The platform features a chat-based interface that supports multi-turn dialogue, allowing students to iteratively refine their prompts. For each message, the full conversation history is provided to the GenAI along with a system instruction to exclude testing code and main function. We employ GPT-4o mini~\cite{GPT4omini} as the assistant model, due to its popularity and cost-effectiveness. Students can reset the conversation any time using the ``Reset'' button and are required to do so upon reaching the $20$-message limit. Students can execute generated code using the ``Run code'' button, which evaluates it against hidden test cases. These features reflect authentic programming workflows and encourage students to reflect on code behavior, reason through errors, and iteratively improve their solutions.

A key feature of the platform is the code editing functionality, as Figure~\ref{fig.illustration} illustrates. For code generated in the most recent GenAI response, students can select ``Edit code'', which loads it into a dedicated editor, above the message input area. Here, students can review and revise chosen code manually. After editing, students can execute their modified code using the ``Run code'' button present in the editor, receiving feedback from hidden test cases. Optionally, they can attach edited code to a new natural-language message and press ``Send'', thus continuing the conversation with the GenAI. By supporting smooth transitions between natural-language prompting and manual code editing, the platform empowers students to take greater ownership of their solutions and provides rich data for understanding how they navigate AI-assisted programming.

\begin{figure*}[t!]
    \centering
    \begin{subfigure}[t]{0.3\linewidth}
        \centering
        \includegraphics[width=\linewidth,trim={0.5cm 0cm 0.5cm 0cm},clip]{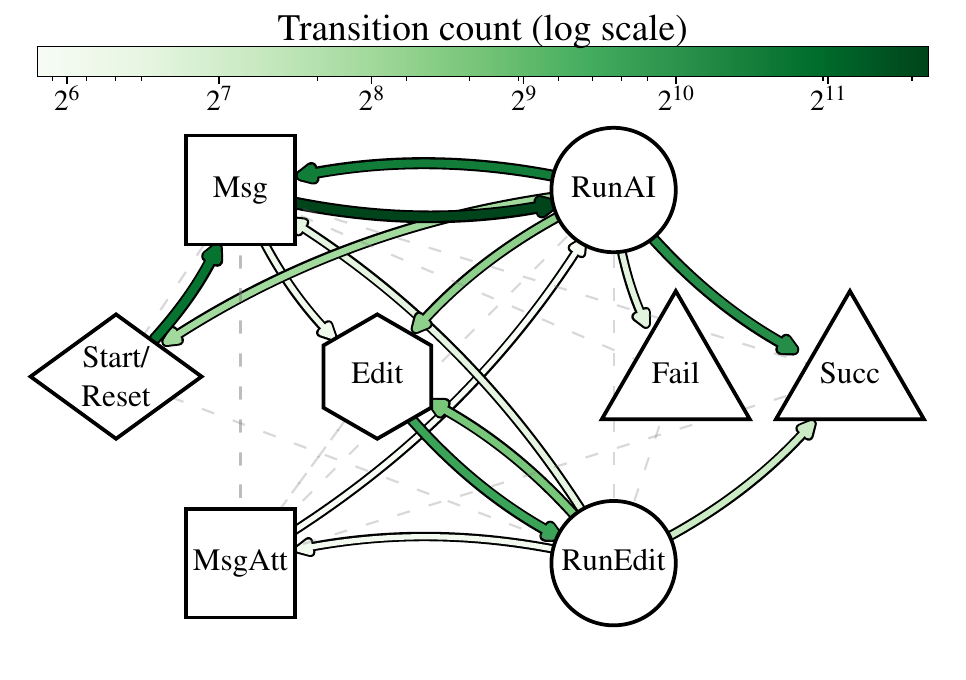}
        \caption{Transition counts between states.}
        \label{fig.rq1.markov}
    \end{subfigure}
    \hfill
    \begin{subfigure}[t]{0.69\linewidth}
        \centering
        \includegraphics[width=\linewidth,trim={0 0 0.4cm 0},clip]{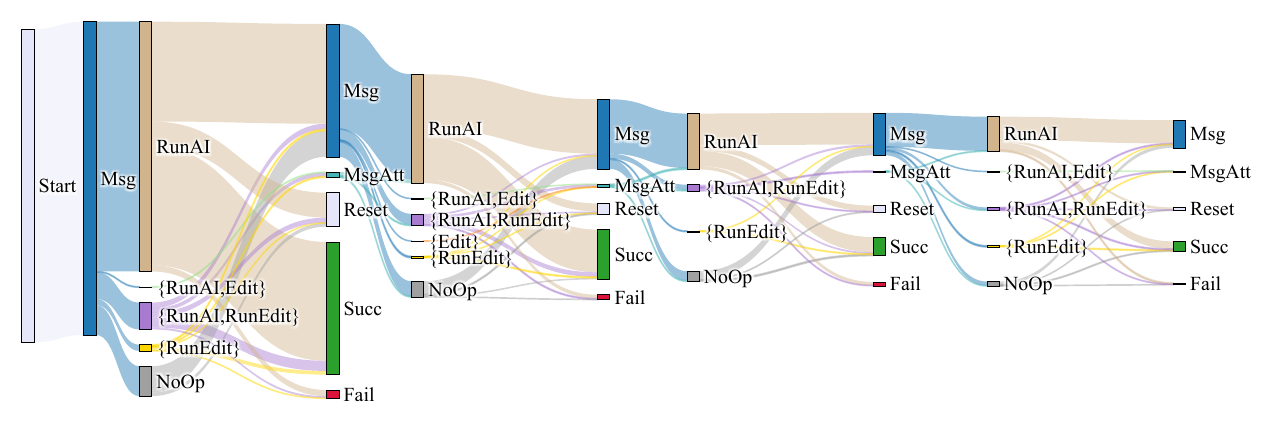}
        \caption{Flow diagram capturing aggregate per-turn patterns.}
        \label{fig.rq1.sankey}
    \end{subfigure}
    \caption{\looseness-1RQ1. Patterns of prompting and editing across problems. (a) presents aggregate transitions between interaction states: Start/Reset, NL message (Msg), NL message with attached code-edit (MsgAtt), execution of GenAI-generated code (RunAI), manual edit (Edit), execution of edited code (RunEdit), and the terminal outcomes Success (Succ) or Failure (Fail). (b) shows sessions structured into alternating odd turns with messaging/terminal events (Msg, MsgAtt, Reset, Succ, Fail), and even turns summarizing execution/edit actions: RunAI (execution of GenAI code only); \{Edit\} (potentially multiple edits without execution); \{RunAI, Edit\} (execution of GenAI code followed by edits without execution); \{RunEdit\} (potentially multiple loops of editing and running edited code); \{RunAI, RunEdit\} (both  execution of GenAI code and edit-execute loops); and NoOp (no execution or edits).
    }
    \Description[Two diagrams summarizing prompting and editing behavior across sessions]{Two-panel figure summarizing aggregate patterns of prompting and editing across student sessions. The left panel is a transition graph whose nodes represent interaction states such as Start or Reset, message, message with attached edited code, run generated code, edit, run edited code, fail, and success. The thickest paths show frequent looping between messaging and running generated code, with another common path moving from failed execution into edit-and-run behavior. The right panel is a turn-by-turn flow diagram over the first nine turns of a session. It shows that many sessions reach success within the first few turns, and that manual editing most often appears after running generated code rather than in isolation.}
    \label{fig.rq1}
\end{figure*}

\subsection{Study Setup}

\looseness-1To understand how students integrate natural-language prompting and manual code editing, we used the platform as part of a laboratory activity in an introductory Python programming course, taught at the University of Auckland. To help our investigation, we collected student queries, GenAI responses, code executions, and code edits. We logged only edits that are executed or attached to a new prompt, ensuring that logged changes represent meaningful engagement. Students were also invited to provide open-ended reflections after finishing all the programming tasks. Specifically, students were asked to comment on what they found to be the ideal combination of skills in natural-language prompting and manual code editing when solving these tasks, as well as to describe the strategies they used, including how they decided when to prompt or edit and what factors influenced their choices. As part of the course, students completed three invigilated, computer-based code writing tests during the semester. These tests took place without access to a GenAI model and required students to write and debug code from scratch.

\looseness-1The laboratory activity was conducted over a three-day period towards the end of the semester. Students had access to $9$ programming problems, distributed across three difficulty categories and adapted to Python from a set of Prompt Problems previously used in literature for introductory programming~\cite{padurean2025prompt}. The Basic Functions category included three single-function problems, referred to as P1, P2, and P3 in our later analysis: counting negatives in a list, summing even numbers in a list, and finding the index of the last zero in a list, respectively. The Advanced Functions category featured more challenging single-function tasks, such as sorting a sublist within specified indices, updating a matrix by propagating $1$s across rows and columns, and adding binary numbers represented as lists. The Classes \& Functions category comprised multi-function problems requiring helper functions and the use of classes or data structures, including password validation, guiding a robot's movement on a grid, and checking if an expression's parentheses are balanced using a stack. The problems in the first category (P1, P2, and P3) were mandatory as part of the laboratory assignment, while the latter two categories were optional for students seeking additional practice.

\subsection{Descriptive Statistics and Analysis}

\looseness-1A total of $355$ students engaged with the tasks. $354$ of them attempted the Basic Functions problems, with $318$ ($89.6\%$) solving every problem. Additionally, $135$ students ($38.0\%$) attempted at least one Advanced Functions problem, and $53$ ($14.9\%$) at least one from the Classes \& Functions category; $52$ students ($14.6\%$) attempted problems from all categories. Among the students, $330$ completed all three tests throughout the semester. To analyze competence effects, we split students based on average test scores: $180$ scored $\geq8$ and $150$ scored $<8$. The median was $7.67$, which we rounded to the nearest integer (i.e., $8$), yielding two groups of comparable size.

To analyze differences between problems of different difficulties and students with different competences, we conducted pairwise Mann-Whitney U tests~\citep{mann1947test} to see if the distributions are different.
%%%%%%%%%%%%%%%%%%%%%%%%%%%%%%%%%%%%%%%%%%%%%%%%%%%%%%%%%%%%%%%%%%
%%%%%%%%%%%%%%%%%%%%%%%%%%%%%%%%%%%%%%%%%%%%%%%%%%%%%%%%%%%%%%%%%%

%%%%%%%%%%%%%%%%%%%%%%%%%%%%%%%%%%%%%%%%%%%% RESULTS
%%%%%%%%%%%%%%%%%%%%%%%%%%%%%%%%%%%%%%%%%%%%%%%%%%%%%%%%%%%%%%%%%%
%%%%%%%%%%%%%%%%%%%%%%%%%%%%%%%%%%%%%%%%%%%%%%%%%%%%%%%%%%%%%%%%%%
\begingroup
  \captionsetup[subfigure]{justification=centering,singlelinecheck=false}

\begin{figure*}[t!]
    \centering

    \vspace{6mm}
    \raisebox{-1mm}[0pt][0pt]{%
      \makebox[0.50\linewidth][r]{%
        \includegraphics[width=0.465\linewidth]{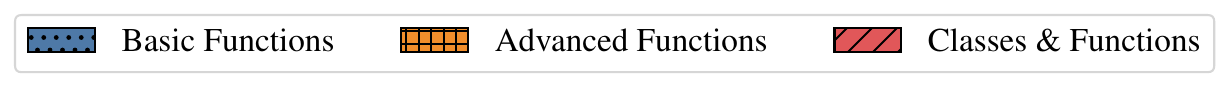}
      }
    }%
    \raisebox{-1mm}[0pt][0pt]{%
      \makebox[0.495\linewidth][r]{%
        \includegraphics[width=0.465\linewidth]{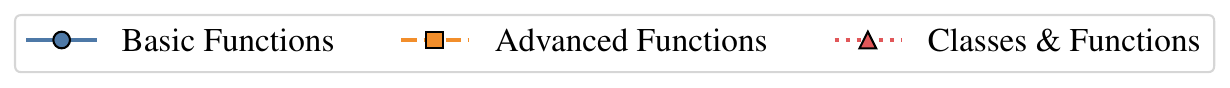}
      }
    }

    \begin{subfigure}[t]{0.245\linewidth}
        \centering
        \includegraphics[width=0.99\linewidth]{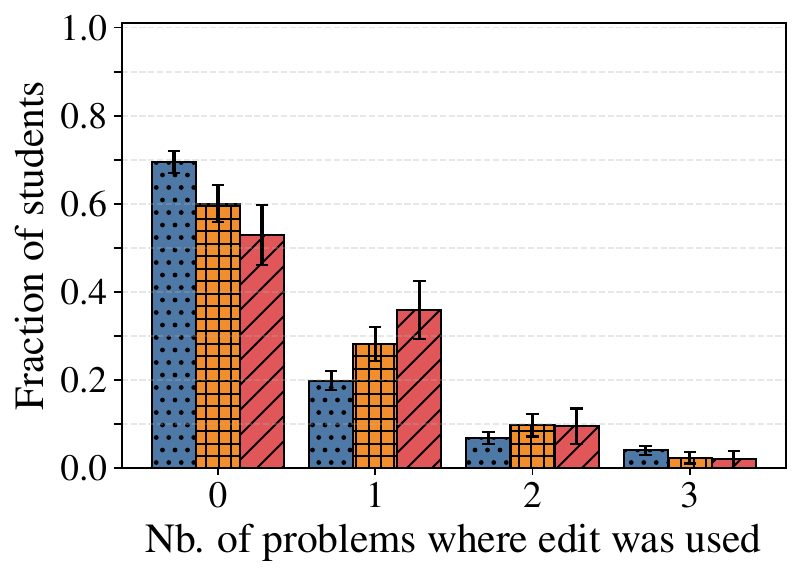}
        \vspace{-5mm}
        \caption{Students editing $k$ problems.}
        \label{fig.rq2.category.problems}
    \end{subfigure}%
    \hfill
    \begin{subfigure}[t]{0.245\linewidth}
        \centering
        \includegraphics[width=0.99\linewidth]{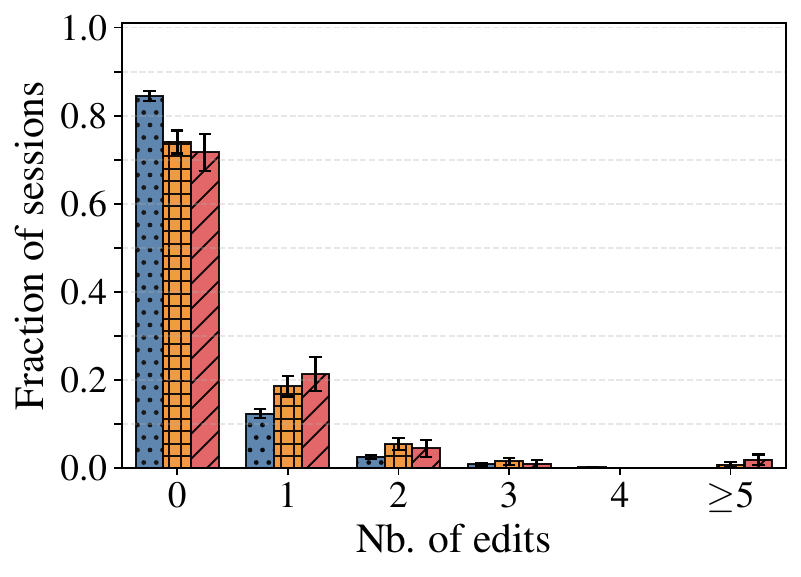}
        \vspace{-5mm}
        \caption{Number of edits per session.}
        \label{fig.rq2.category.edits}
    \end{subfigure}%
    \hfill
    \begin{subfigure}[t]{0.245\linewidth}
        \centering
        \includegraphics[width=0.99\linewidth]{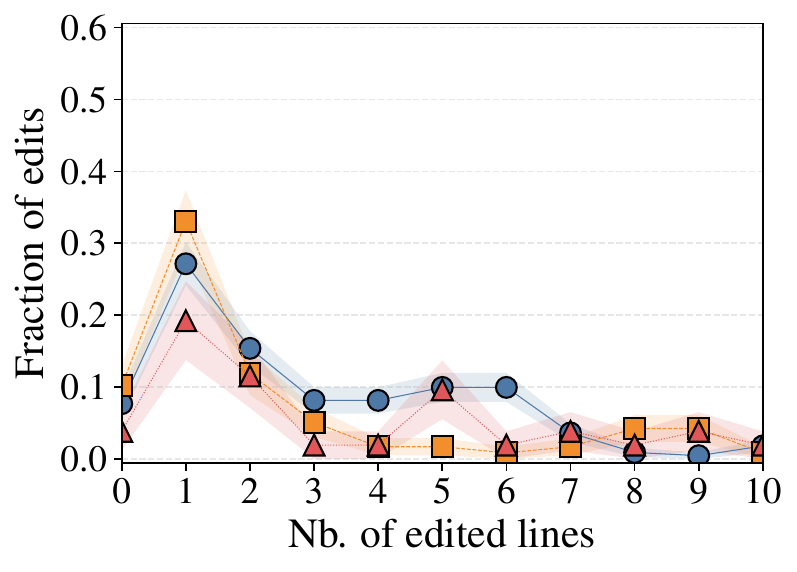}
        \vspace{-5mm}
        \caption{Lines changed per edit.}
        \label{fig.rq2.category.lines}
    \end{subfigure}%
    \hfill
    \begin{subfigure}[t]{0.245\linewidth}
        \centering
        \includegraphics[width=0.99\linewidth]{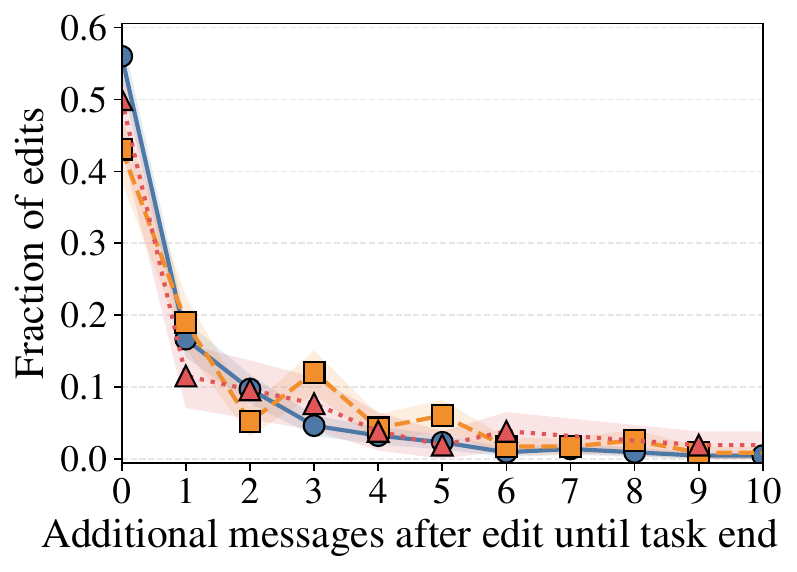}
        \vspace{-5mm}
        \caption{Additional messages after edit.}
        \label{fig.rq2.category.actions}
    \end{subfigure}%
    \caption{RQ2. Editing activity by problem category. (a) Fraction of students who edited code in $k$ of the problems they attempted. (b) Distribution of manual edits per session (i.e., student attempting a problem, possibly across multiple conversations). (c) Distribution of lines changed per retained edit (line-level Levenshtein distance); $0$ indicates code executed or attached unchanged or reverted by the end of the edit cycle. (d) Number of additional messages sent after an edit before reaching a terminal outcome.}
    \Description[Four plots comparing editing behavior across problem categories]{Four plots compare editing activity across three problem categories: Basic Functions, Advanced Functions, and Classes and Functions. The first plot shows how many attempted problems involved any editing, with editing used more often in the harder categories. The second plot shows the number of edits per session, again indicating more editing in the harder categories, although many sessions still contain no edits. The third plot shows lines changed per edit and peaks at single-line edits for all categories. The fourth plot shows how many additional messages follow an edit before the session ends, with most edits followed by only a small number of further interactions. Overall, the figure shows that editing becomes more common on harder tasks, but edits remain brief and localized.}
    \label{fig.rq2}
\end{figure*}
\endgroup

\section{Results}\label{sec.results}

In this section, we present results for each research question (RQ).

\subsection{RQ1: Interleaving Prompts and Code Edits}

\looseness-1To gain a general understanding of how students interleave natural language prompting and manual code editing, we analyze their behavioral patterns from interaction logs (see Figure~\ref{fig.rq1}). We then complement this with students' perceptions from reflection responses. 

\looseness-1Figure~\ref{fig.rq1.markov} presents the most frequent transitions between interaction states, aggregated across all sessions (i.e., student attempting a problem). For each session, we ordered states chronologically and counted all adjacent state pairs, then selected the top $15$ transitions by frequency to highlight the most important patterns. The figure shows that the most common interaction pattern is looping between messaging and running GenAI-generated code, with a notable subset transitioning directly from RunAI to success. At the same time, there is a substantial pathway from unsuccessful execution of generated code into the Edit-RunEdit loop, indicating that many students refine solutions through manual edits and check their correctness by running them as well. Overall, the diagram suggests that both prompting and code editing are actively interleaved in students' problem-solving workflows, though their relative use varies.

Figure~\ref{fig.rq1.sankey} provides a turn-based flow diagram of the first nine turns per session\footnote{In the figure, we aggregate data across all problem categories, since the diagrams show the same overall pattern across task difficulty levels.}, clarifying how students interleave prompting and editing actions over time. The diagram alternates between odd-numbered turns, which represent messaging or terminal events, and even-numbered turns, which group executions and edit actions into sets. Three insights stand out from the visualization. First, a substantial number of sessions reach a solution within the first few turns, reflecting that initial prompts can often be effective. Second, manual editing typically appears as a reaction to unsuccessful executions of GenAI-generated code, as indicated by the thicker flow through the combined state \{RunAI, RunEdit\}, compared to the narrower \{RunEdit\}-only path. This confirms students mostly introduce manual edits after testing GenAI code. Finally, the ratio between the \{RunAI\}-only and \{RunAI, RunEdit\} flows remains relatively stable across turns, indicating that students introduce edits at a wide range of points during their sessions.

\looseness-1To complement these behavioral patterns, we analyzed responses to a Likert-style item asking what combination of prompting and editing they found most effective. They chose among five options (student share shown in brackets): 1. ``Only natural language prompting'' (5.36\%); 2. ``Mostly natural language prompting with a little manual code editing'' (24.67\%); 3. ``An equal mix of natural language prompting and manual code editing'' (44.77\%); 4. ``Mostly manual code editing with a little natural language prompting'' (22.52\%); 5. ``Only manual code editing'' (2.68\%). Responses followed a normal distribution centered on an equal mix of prompting and editing, suggesting that students view the two approaches as complementary.

\looseness-1Furthermore, students responded to an open-ended question describing how they chose between prompting and editing. We found three key themes: prompting for speed, editing for precision, and learning with AI. Many students emphasized the efficiency of using prompting to quickly generate initial solutions, turning to manual edits for small, as one student noted: \emph{``I used AI prompts first to get the basic code quickly. When I needed small fixes, I edited the code myself.''} Others described a hybrid strategy in which AI-generated code served as a starting point that they incrementally corrected or refined through manual editing to ensure accuracy and handle edge cases. 
Finally, several students prioritized learning, using prompting to explore unfamiliar concepts and editing to consolidate understanding and retain control over the solution. As one student reflected, \emph{``If it is a question I know how to do, then I use manual code, otherwise I use a prompt and learn from the AI generated code.''}

\rqbox{\textbf{Answer to RQ1:}
Students tend towards using prompts to generate initial solutions and applying edits after testing GenAI code. They view prompting as enabling speed, exploration, and learning, while editing supports precision, verification, and control.}

%%%%%%%%%
\subsection{RQ2: Effects of Task Complexity}

\begingroup
  \captionsetup[subfigure]{justification=centering,singlelinecheck=false}

\begin{figure*}[t!]
    \centering
  \begin{subfigure}[t]{0.32\linewidth}
    \centering
    \includegraphics[width=\linewidth]{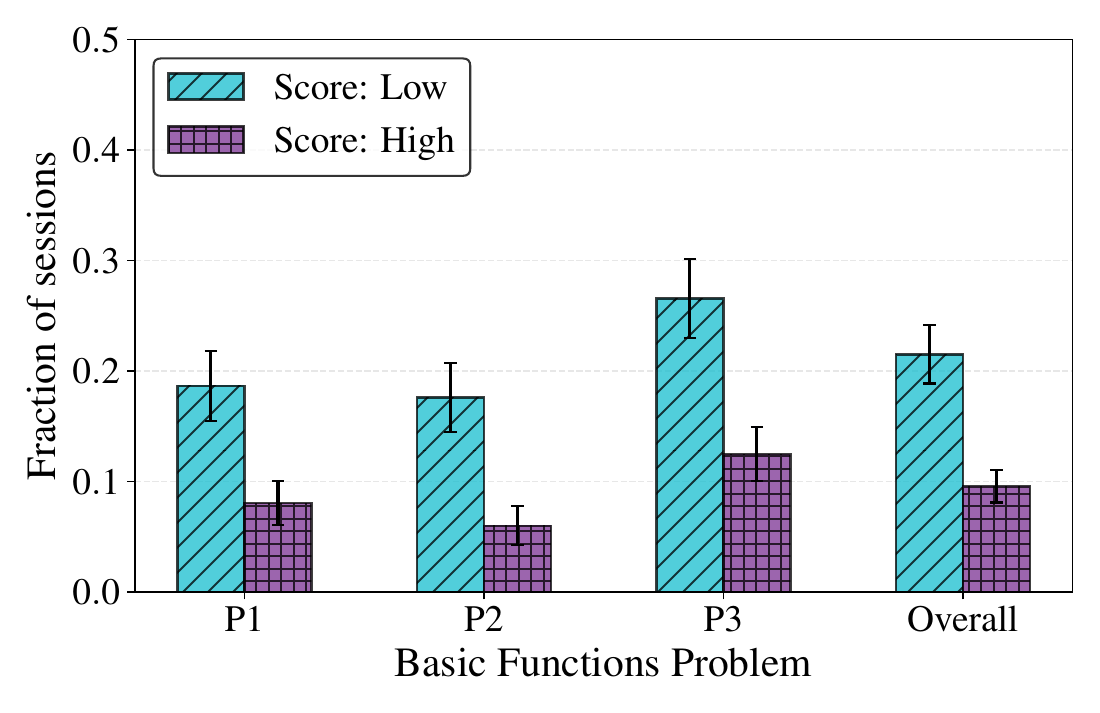}
    \vspace{-6mm}
    \caption{Proportion of sessions with Edit}
    \label{fig.rq3.cohort.edit}
  \end{subfigure}%
  \hfill
  \begin{subfigure}[t]{0.32\linewidth}
    \centering
    \includegraphics[width=\linewidth]{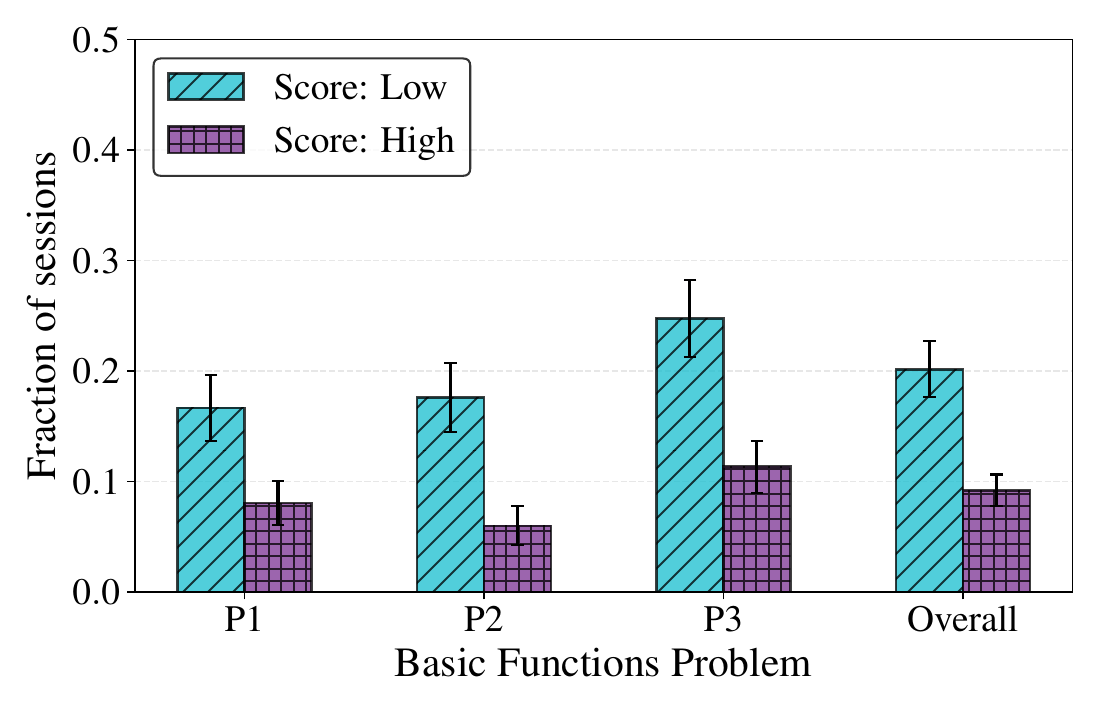}
    \vspace{-6mm}
    \caption{Proportion of sessions with RunEdit}
    \label{fig.rq3.cohort.runedit}
  \end{subfigure}%
  \hfill
  \begin{subfigure}[t]{0.32\linewidth}
    \centering
    \includegraphics[width=\linewidth]{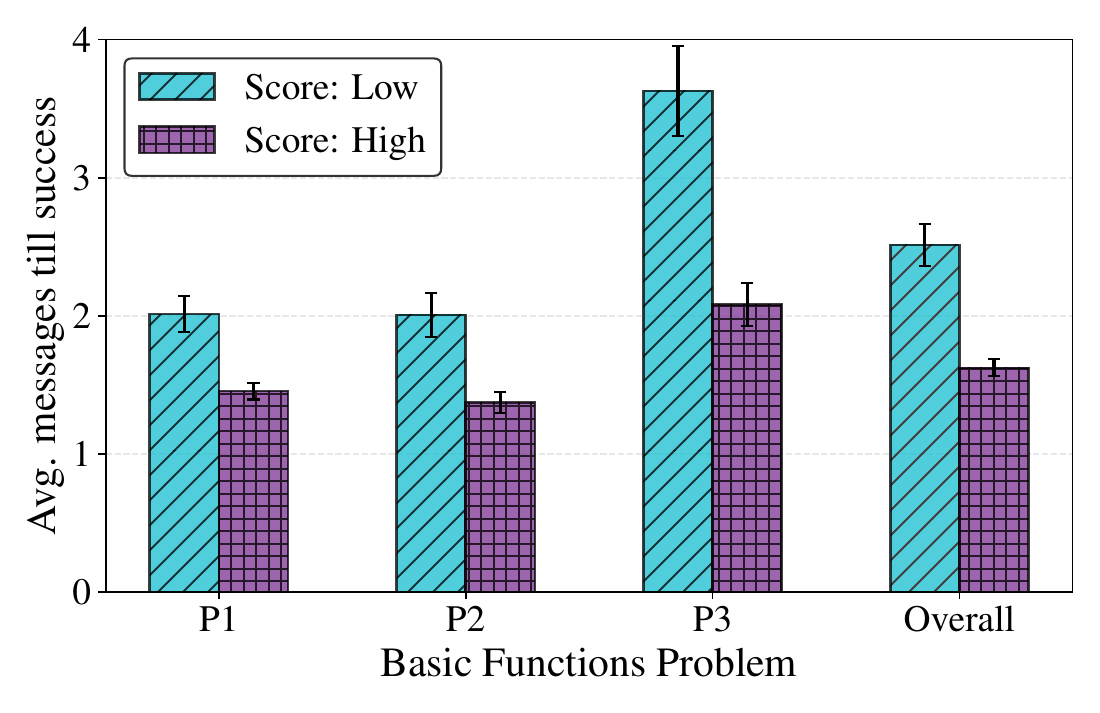}
    \vspace{-6mm}
    \caption{Mean messages till first success}
    \label{fig.rq3.cohort.msg}
  \end{subfigure}%
        \caption{\looseness-1RQ3. Activity by group (\emph{LowScore}, \emph{HighScore}) on Basic Functions problems: P1, P2, P3, and Overall (averaged across the three problems). (a) Proportion of sessions that include at least one Edit action. (b) Proportion of sessions that include at least one RunEdit action. (c) Mean number of messages sent until the first successful attempt (absolute prompting volume).}
        \Description[Three bar charts comparing lower- and higher-scoring students on editing behavior]{Three bar charts compare two student groups, LowScore and HighScore, on the three Basic Functions problems and on an overall average. The first chart shows the proportion of sessions containing at least one Edit action, and the second shows the proportion containing at least one RunEdit action. In both charts, the HighScore group has lower bars than the LowScore group across all problems. The third chart shows the mean number of messages until first success, where the HighScore group again has lower values. Overall, the figure shows that higher-performing students used less manual editing and reached success with fewer interactions.}
    \label{fig.rq3}
\end{figure*}
\endgroup

\looseness-1To study how editing behavior varies with task difficulty, we analyze interaction logs across problem categories. For our analysis, we use the final code version from each Edit-RunEdit loop and look at four indicators of editing behavior in Figure~\ref{fig.rq2}.

\looseness-1Figures~\ref{fig.rq2.category.problems}~and~\ref{fig.rq2.category.edits} show that editing was more common when students worked on more complex problems.
In Figure~\ref{fig.rq2.category.problems}, only about $31\%$ of students edited at least one Basic Functions problem, compared to roughly $40\%$ in Advanced Functions and $47\%$ in Classes \& Functions. Figure~\ref{fig.rq2.category.edits} shows the same trend at the session level, with only about $16\%$ of Basic Functions sessions containing any edit, compared to around $26\%$ in Advanced Functions and $28\%$ in Classes \& Functions (Basic vs.\ Advanced: $U = 128834.5, p < 0.001$; Basic vs.\ Classes \& Functions: $U = 50514, p < 0.001$).\footnote{For pairwise comparisons between the three problem categories, we applied Bonferroni correction~\citep{dunn1959estimation} with a factor of $3$.} Because Advanced Functions and Classes \& Functions were optional, we repeated the analysis on the $52$ students who attempted at least one problem in each category (visualizations not shown due to space constraints). The overall trends remained consistent.

\looseness-1Figure~\ref{fig.rq2.category.lines} shows that edit scope remains stable across categories. Most edit cycles involved a single-line change and were typically followed by few additional interaction steps before the session finished (Figure~\ref{fig.rq2.category.actions}). We refer to this combination of localized edits near the end of the session as \emph{last-mile repair}. This scope did not significantly differ by task difficulty. Together, these patterns suggest that regardless of problem complexity, students primarily used editing for small corrective adjustments and not extensive rewrites.

\rqbox{\textbf{Answer to RQ2:}
Editing was more common on harder tasks, yet edits stayed brief and local rather than involving major rewrites.}

%%%%%%%%
\subsection{RQ3: Effects of Programming Competence}
To understand how programming competence relates to editing behavior, we analyze Basic Functions, the only mandatory category. This choice limits confounding effects from task difficulty and optional participation. We measure competence using each student's mean score across the three invigilated code-writing tests used for course assessment (see Section~\ref{sec.methodology}). We split students into two groups, the \emph{HighScore} group (mean $\geq 8$) and the \emph{LowScore} group (mean $< 8$), with Figure~\ref{fig.rq3} comparing their code-editing behavior.

\looseness-1Figures~\ref{fig.rq3.cohort.edit}~and~\ref{fig.rq3.cohort.runedit} show that students in the \emph{HighScore} group were less likely to use manual editing.
In Figure~\ref{fig.rq3.cohort.edit}, across all problems (i.e., Overall in the figure), only about $9.6\%$ of sessions from the \emph{HighScore} group contained any Edit action, compared to roughly $21.5\%$ of sessions from the \emph{LowScore} group ($U = 10984, p < 0.001$).
Figure~\ref{fig.rq3.cohort.runedit} shows the same pattern for RunEdit action, with about $9.2\%$ of sessions from the \emph{HighScore} group containing any RunEdit action across all problems (i.e., Overall in the figure), compared to around $20.2\%$ for the \emph{LowScore} group ($U = 11169.5, p < 0.001$).
We then examined whether this difference is specific to editing or reflects lower overall interaction.
Figure~\ref{fig.rq3.cohort.msg} shows that students in the \emph{HighScore} group solved the problems quicker, requiring significantly fewer messages per session before success ($U = 8975.5, p < 0.001$).

\looseness-1Together, these results suggest that higher-performing students reached correct solutions with fewer iterations across the full workflow, but that they also entered the edit-and-run loop less often. We believe that this happens because higher-performing students were better able to reason through the problems and formulate more effective prompts early on, which may have reduced the need for later manual edits after unsuccessful runs. Prior work indicates that effective GenAI use relies on metacognitive skills such as planning and monitoring one's problem-solving process \cite{tankelevitch2024metacognitive}, and our findings are consistent with evidence that students with higher scores and confidence may use GenAI more effectively and experience fewer difficulties \cite{prather2024widening}. However, because we do not analyze prompt quality directly, this interpretation remains speculative.

\rqbox{\textbf{Answer to RQ3:}
More proficient students edited less frequently and required less overall interaction to achieve success.}

%%%%%%%%
\subsection{Limitations}

We acknowledge several limitations of our study. First, only the Basic Functions problems were mandatory, while more complex tasks were optional, introducing potential self-selection bias. It would be valuable to conduct an investigation when all students complete tasks of varying difficulty. Second, we did not conduct an A/B study to compare prompting-only and prompting-plus-editing conditions in a controlled way. A randomized study where editing capabilities are selectively enabled could more directly measure the benefits of code editing. Third, lacking demographic data and prior GenAI experience limited our possible analyses. We also did not analyze prompt content directly, which limits how strongly we can interpret differences in students' strategies. Finally, our study used a custom-built platform that tightly integrates chat-based prompting, code execution, code attachment, and direct code editing. While the overall pattern of prompting, testing, and localized repair may extend to other programming environments (e.g., other platforms or IDEs), the ease and timing of these transitions may depend on interface features specific to the platform in our study. 
%%%%%%%%%%%%%%%%%%%%%%%%%%%%%%%%%%%%%%%%%%%%%%%%%%%%%%%%%%%%%%%%%%
%%%%%%%%%%%%%%%%%%%%%%%%%%%%%%%%%%%%%%%%%%%%%%%%%%%%%%%%%%%%%%%%%%

%%%%%%%%%%%%%%%%%%%%%%%%%%%%%%%%%%%%%%%%%%%% CONCLUSION
%%%%%%%%%%%%%%%%%%%%%%%%%%%%%%%%%%%%%%%%%%%%%%%%%%%%%%%%%%%%%%%%%%
%%%%%%%%%%%%%%%%%%%%%%%%%%%%%%%%%%%%%%%%%%%%%%%%%%%%%%%%%%%%%%%%%%

\section{Conclusion}\label{sec.conclusion}

\looseness-1Our study investigated a key question in AI-assisted programming: when do students revise their prompt, and when do they dive into the code? We found that students combined prompting and manual editing in their workflow, typically introducing edits after a generated solution failed execution. These manual edits were usually concise, serving as targeted, last-mile corrections. Student reflections highlighted prompting for efficiency, editing for precision, and both as valuable tools for learning. Edits became more common as tasks increased in complexity, while higher-performing students relied less on manual editing and reached success with fewer overall interactions. These findings emphasize the need to support both skills in instruction, teaching students how to decide when to re-prompt versus apply an edit after testing. Tools and teaching materials can also offer scaffolding around the edit-run loop when students struggle. Developing this flexibility requires a solid understanding of the problem and the underlying programming concepts.

\looseness-1Our work opens up directions for future research on how students learn to balance prompting and editing. First, studies could introduce controlled bugs into otherwise correct completions to cultivate students' ability to detect and correct subtle issues. Second, research might investigate how students make strategic decisions between prompting and editing by presenting tasks that require them to justify which approach would be more effective. Third, longer-term studies could track how editing behaviors evolve in open-ended domains like data analysis, visualization, or machine learning, where last-mile fixes may become broader refactors. These directions could help students develop both precise prompting strategies and fluency in concise, high-impact edits.

%%%%%%%%%%%%%%%%%%%%%%%%%%%%%%%%%%%%%%%%%%%%%%%%%%%%%%%%%%%%%%%%%%
%%%%%%%%%%%%%%%%%%%%%%%%%%%%%%%%%%%%%%%%%%%%%%%%%%%%%%%%%%%%%%%%%%

%%%%%%%%%%%%%%%%%%%%%%%%%%%%%%%%%%%%%%%%%%%% ACKNOWLEDGEMENTS
%%%%%%%%%%%%%%%%%%%%%%%%%%%%%%%%%%%%%%%%%%%%%%%%%%%%%%%%%%%%%%%%%%
%%%%%%%%%%%%%%%%%%%%%%%%%%%%%%%%%%%%%%%%%%%%%%%%%%%%%%%%%%%%%%%%%%
\begin{acks}
\looseness-1Procedures used in the study involve analysis of de-identified archival student work and have been approved by the University of Auckland Human Ethics Committee under reference nb. UAHPEC25279. This work was supported by Research Council of Finland grant \#356114. Funded/Cofunded by the European Union (ERC, TOPS, 101039090). Views and opinions expressed are however those of the author(s) only and do not necessarily reflect those of the European Union or the European Research Council. Neither the European Union nor the granting authority can be held responsible for them.
\end{acks}

%%%%%%%%%%%%%%%%%%%%%%%%%%%%%%%%%%%%%%%%%%%%%%%%%%%%%%%%%%%%%%%%%%
%%%%%%%%%%%%%%%%%%%%%%%%%%%%%%%%%%%%%%%%%%%%%%%%%%%%%%%%%%%%%%%%%%

%%%%%%%%%%%%%%%%%%%%%%%%%%%%%%%%%%%%%%%%%%%% BIBLIOGRAPHY
%%%%%%%%%%%%%%%%%%%%%%%%%%%%%%%%%%%%%%%%%%%%%%%%%%%%%%%%%%%%%%%%%%
%%%%%%%%%%%%%%%%%%%%%%%%%%%%%%%%%%%%%%%%%%%%%%%%%%%%%%%%%%%%%%%%%%
\bibliographystyle{ACM-Reference-Format}
\balance
\bibliography{main}

%%%%%%%%%%%%%%%%%%%%%%%%%%%%%%%%%%%
%%%%%%%%%%%%%%%%%%%%%%%%%%%%%%%%%%%
\end{document}